\documentclass[aps,reprint,notitlepage,floatfix,superscriptaddress]{revtex4-1}

\usepackage[pdftex]{graphicx}
\usepackage{amsmath,amssymb,amstext}
\usepackage{mathrsfs}
\usepackage{multirow}
\usepackage{comment}
\usepackage{bbold}
\usepackage{dsfont}
\usepackage{color}
\usepackage{hyperref}
\usepackage{tikz}
\usepackage{braket}
\usepackage[colorinlistoftodos]{todonotes}
\usepackage{soul}

\renewcommand{\eqref}[1]{Eq.~(\ref{#1})}
\newcommand{\figref}[1]{Fig.~\ref{#1}}
\newcommand{\tabref}[1]{Tab.~\ref{#1}}

\begin{document}
\title{Talbot Effect of orbital angular momentum lattices with single photons}

\author{S. Schwarz}
 \email{sacha.schwarz@uwaterloo.ca}
\affiliation{Institute for Quantum Computing, University of Waterloo, Waterloo,
  ON, Canada, N2L 3G1} 
\affiliation{Department of Physics \& Astronomy, University of Waterloo,
  Waterloo, ON, Canada, N2L 3G1}
       \author{C. Kapahi}
  \affiliation{Institute for Quantum Computing, University of Waterloo, Waterloo,
  ON, Canada, N2L 3G1} 
  \affiliation{Department of Physics \& Astronomy, University of Waterloo, Waterloo, ON, Canada, N2L 3G1}
\author{R. Xu}
\affiliation{Institute for Quantum Computing, University of Waterloo, Waterloo, ON, Canada, N2L 3G1} 
\affiliation{Department of Electrical \& Computer Engineering, University of Waterloo, Waterloo, ON, Canada, N2L 3G1}
   \author{A. R. Cameron}
\affiliation{Institute for Quantum Computing, University of Waterloo, Waterloo,
  ON, Canada, N2L 3G1} 
\affiliation{Department of Physics \& Astronomy, University of Waterloo,
  Waterloo, ON, Canada, N2L 3G1}
 \author{D. Sarenac}
 \affiliation{Institute for Quantum Computing, University of Waterloo, Waterloo,
 ON, Canada, N2L 3G1} 
  \affiliation{Department of Physics \& Astronomy, University of Waterloo,
  Waterloo, ON, Canada, N2L 3G1}
  \author{J. P. W. MacLean}
  \affiliation{Institute for Quantum Computing, University of Waterloo, Waterloo,
  ON, Canada, N2L 3G1} 
  \affiliation{Department of Physics \& Astronomy, University of Waterloo,
  Waterloo, ON, Canada, N2L 3G1}
          \author{K. B. Kuntz}
  \affiliation{Institute for Quantum Computing, University of Waterloo, Waterloo,
  ON, Canada, N2L 3G1} 
 \affiliation{Department of Physics \& Astronomy, University of Waterloo,
  Waterloo, ON, Canada, N2L 3G1}
    \author{D. G. Cory}
\affiliation{Institute for Quantum Computing, University of Waterloo, Waterloo,
  ON, Canada, N2L 3G1} 
\affiliation{Department of Chemistry, University of Waterloo, Waterloo, ON, Canada, N2L3G1}
\affiliation{Perimeter Institute for Theoretical Physics, Waterloo, ON, Canada, N2L2Y5}
\affiliation{Canadian Institute for Advanced Research, Toronto, ON, Canada, M5G 1Z8}
         \author{T. Jennewein}
  \affiliation{Institute for Quantum Computing, University of Waterloo, Waterloo,
  ON, Canada, N2L 3G1} 
  \affiliation{Department of Physics \& Astronomy, University of Waterloo,
  Waterloo, ON, Canada, N2L 3G1}
            \author{K.J. Resch}
 \affiliation{Institute for Quantum Computing, University of Waterloo, Waterloo,
  ON, Canada, N2L 3G1} 
  \affiliation{Department of Physics \& Astronomy, University of Waterloo,
  Waterloo, ON, Canada, N2L 3G1}
    \author{D.A. Pushin}
     \email{dmitry.pushin@uwaterloo.ca}
  \affiliation{Institute for Quantum Computing, University of Waterloo, Waterloo,
  ON, Canada, N2L 3G1} 
  \affiliation{Department of Physics \& Astronomy, University of Waterloo,
  Waterloo, ON, Canada, N2L 3G1}
  
\begin{abstract}
The self-imaging, or Talbot Effect, that occurs with the propagation of periodically structured waves has enabled several unique applications in optical metrology, image processing, data transmission, and matter-wave interferometry.  In this work, we report on the first demonstration of a Talbot Effect with single photons prepared in a lattice of orbital angular momentum (OAM) states. We observe that upon propagation, the wavefronts of the single photons manifest self-imaging whereby the OAM lattice intensity profile is recovered. Furthermore, we show that the intensity at fractional Talbot distances is indicative of a periodic helical phase structure corresponding to a lattice of OAM states. 
This phenomenon is a powerful addition to the toolbox of orbital angular momentum and spin-orbit techniques that have already enabled many recent developments in quantum optics.
\end{abstract}
\maketitle

The Talbot Effect~\cite{talbot1836lxxvi} is a near-field diffraction phenomenon whereby periodic phase and amplitude modulations are self-imaged due to free-space propagation. In accordance with Fresnel diffraction~\cite{wen2013talbot}, replicas of a periodic transverse intensity profile reappear after a specific propagation distance known as the Talbot length. The Talbot Effect has been demonstrated in numerous areas of research involving linear and nonlinear optical waves \cite{iwanow2005discrete,case2009realization,zhang2010nonlinear}, single photons \cite{song2011experimental,deachapunya2016realization}, x-rays \cite{pfeiffer2008hard}, matter-waves \cite{chapman1995near,deng1999temporal,wu2007demonstration,Pfeiffer2006}, exciton polaritons \cite{gao2016talbot}, and Bose-Einstein condensates \cite{makhalov2019order}. The Talbot Effect has a diverse array of applications in optical metrology \cite{rogers2012super}, imaging processing \cite{wang2009light}, and lithography \cite{isoyan2009talbot,solak2011displacement,liu2019polarization}, with potential in data transmission \cite{azana1999temporal}.

Here we consider the Talbot Effect manifested by lattices of orbital angular momentum (OAM) states. The OAM degree of freedom of light has garnered significant interest in various fields ranging from optical manipulation and high-bandwidth communication \cite{he1995direct,friese1996optical,wang2012terabit,zhang2019coherent} to quantum information processing \cite{Andersen2006,erhard2018twisted}. In addition to the photonic applications, OAM beams have been extended to neutrons \cite{clark2015controlling,sarenac2016holography,Sarenac201906861} and electrons \cite{uchida2010generation,mcmorran2011electron}.

The Talbot Effect has been considered with classical light as well as OAM lattices \cite{courtial2006angular,vyas2007interferometric,wei2009generation,sarenac2018generation,gao2018quasi,Hu2018,Rasouli2019}. In this Letter, we report the first demonstration of the Talbot Effect with single photons prepared in a lattice of OAM states. The extension of the Talbot Effect to single photons and OAM techniques offers the possibility of utilizing quantum information processing protocols, such as remote state preparation, to leverage quantum communication advantages \cite{barreiro2010remote}. Furthermore, self-imaging has potential applications in implementing quantum logic operations as qudits may be encoded in the transverse spatial profile of single photons~\cite{farias2015quantum,sawada2018experimental}. 

\begin{figure*}[ht!]
   \centering
   \includegraphics[width=\textwidth]{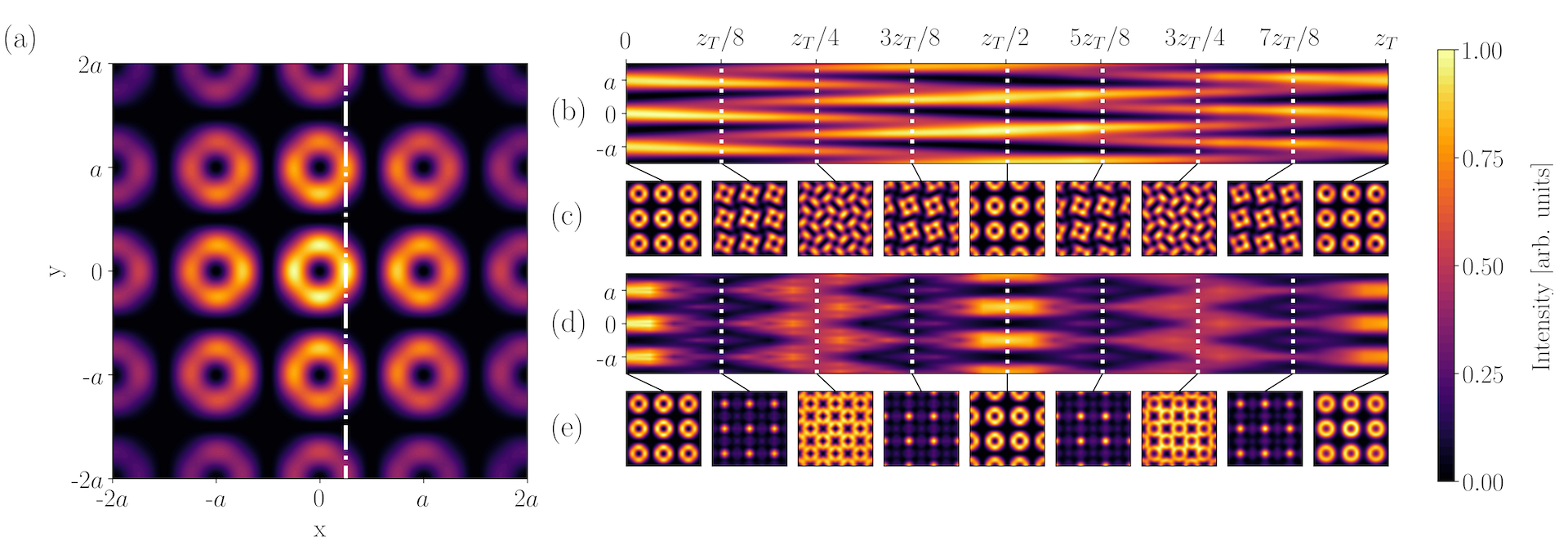}
   \caption{Simulated intensity distributions in both the $xy$ and $yz$ planes, where the beam propagates along $z$. (a) Right-handed circularly polarized light is sent through $N=2$ sets of LOV prism pairs, which yields a lattice of ring-shaped intensity structures when filtered with respect to the left-handed circular polarization, i.e., $I(x,y)=\vert\langle L\vert\ket{\Psi^{N=2}_\text{LOV}}\vert^2$ at propagation distance $z=0$. Note that here the Gaussian intensity envelope $\alpha(x,y)$ with beam waist $w_0=3a$ is added. (b) By plotting the $yz$ intensity at $x=a/4$ (indicated in (a) with the dash dotted white line) we recover what is known as the Talbot carpet. (c) $xy$ intensity patterns at specific propagation distances $z$. (d-e) The Talbot carpet and the $xy$ intensity cross sections when the phase structure of the initial beam is removed. This demonstrates the effect of the OAM lattice phase structure on the intensity profile at the fractional Talbot distances.  
   }
   \label{fig:TalbotTheo}
\end{figure*}

We consider spin-orbit states described by the wavefunction
\begin{equation}
	\ket{\Psi}=
        	A(r,\phi)\left[
            	\cos\left(\frac{ \pi r}{d}\right) \ket{R}
                +ie^{i\ell\phi}\sin\left(\frac{ \pi r}{d}\right)\ket{L}
            \right],
	\label{Eqn:psiSO}
\end{equation}
where $(r,\phi)$ are the cylindrical coordinates, $\ell$ specifies the OAM number, $d$ is the distance in which the polarization state performs a full rotation on the Poincar\'e sphere, $\ket{R}$ and $\ket{L}$ denote the right and left circular polarization states, and $A(r,\phi)$ denotes the envelope. A lattice of spin-orbit states can be obtained by passing circularly polarized light through pairs of birefringent linear gradients whose optical axes are perpendicular to each other~\cite{sarenac2018generation,sarenac2018methods}.  This may be derived by considering the Suzuki-Trotter expansion of the operator $\hat{U}$ which generates the spin-orbit state of \eqref{Eqn:psiSO} when acting on an input state $\ket{R}$, where
\begin{align}
    \hat{U} = e^{ i\frac{\pi}{d}[x\hat{\sigma}_x + y\hat{\sigma}_y] }
    = \lim_{N\to\infty} ( e^{ i\frac{\pi x}{N d} \hat{\sigma}_x } e^{ i \frac{\pi y}{N d} \hat{\sigma}_y } )^N.
    \label{eq:expansion}
\end{align}
Truncating the expansion to $N$ terms, the operators in \eqref{eq:expansion} can be realized by sets of perpendicular birefringent gradients with the general form

\begin{equation} 
    \hat{U}_x=e^{i\frac{\pi}{a}(x-x_0)\hat{\sigma}_x},
    \quad
    \hat{U}_y=e^{i\frac{\pi}{a}(y-y_0)\hat{\sigma}_y},
    \label{eq:shiftOperators}
\end{equation}
    where the origin of the gradients is given by $(x_0, y_0)$, $\hat{\sigma}_{x,y}$ are Pauli matrices, and where $a=Nd$. It was shown in Ref.~\cite{sarenac2018generation} that linear gradients  of Eq.~(\ref{eq:shiftOperators}) may be implemented via ``Lattice of Optical Vorticies'' (LOV) prism pairs. A LOV prism pair consists of two wedge-shaped birefringent prisms where the optical axis of the first prism is along the wedge incline direction and that of the second is offset by $45^\circ$~\cite{sarenac2018generation}. By sending a photon in the right circular polarization state $\ket{R}$ through $N$ sets of LOV prism pairs, we prepare the state
\begin{align}
    \ket{\Psi^{N}_\text{LOV}}= \alpha(x,y)(\hat{U}_x\hat{U}_y)^N \ket{R},
   \label{eq:LOV_state}
\end{align}
where $\alpha(x,y)$ describes the incoming Gaussian beam envelope with beam waist $w_0$. The periodic nature of polarization rotations enables the linear gradients to prepare a two-dimensional lattice of spin-orbit states. 
 
 Filtering on one circular polarization state prepares a periodically structured intensity distribution with a lattice spacing of $a = \lambda (\Delta n \tan(\theta))^{-1}$, where $\Delta n$ and $\theta$ are the birefringence  and the incline angle of the LOV prism pairs respectively, and $\lambda$ is the wavelength. In our experiment we use $N=2$ LOV prism pairs and we filter on  $\vert L \rangle$ to obtain an initial intensity distribution of the form
\begin{align}\nonumber
I(x,y)&=\vert\langle L\vert\ket{\Psi^{N=2}_\text{LOV}}\vert^2\\\nonumber
&=\vert\alpha(x,y)\vert^2\cos^2\left[\frac{ \pi x}{a}\right]\cos^2\left[\frac{ \pi y}{a}\right]\\
&\phantom{=}\times(2-\cos\left[\frac{2 \pi (x+y)}{a}\right]-\cos\left[\frac{2 \pi (x-y)}{a}\right]),
\label{eq:lattice}
\end{align}
\noindent which is depicted in \figref{fig:TalbotTheo}(a). This periodic beam structure imprinted by the LOV prism pairs sets up conditions required for the Talbot Effect. The transmitted light interferes in such a way that after a distance $z_T=2 a^2/\lambda$, the initial periodic intensity pattern reappears. The same intensity distribution also appears at half the distance, $z_T/2$, but with spatial shifts $\Delta a = a/2$ along the $x$- and $y$-directions. Furthermore, the intensity distribution at propagation distances much larger than the Talbot distance  results in the Fourier transform of the initial periodic pattern. The Fraunhofer distance (far-field) is given by  $z_F=8  w_0^2/\lambda$, where $w_0$ is the beam waist. In our setup  $z_F \approx 166$~m.

\begin{figure*}[ht!]
   \centering
   \includegraphics[width=\textwidth]{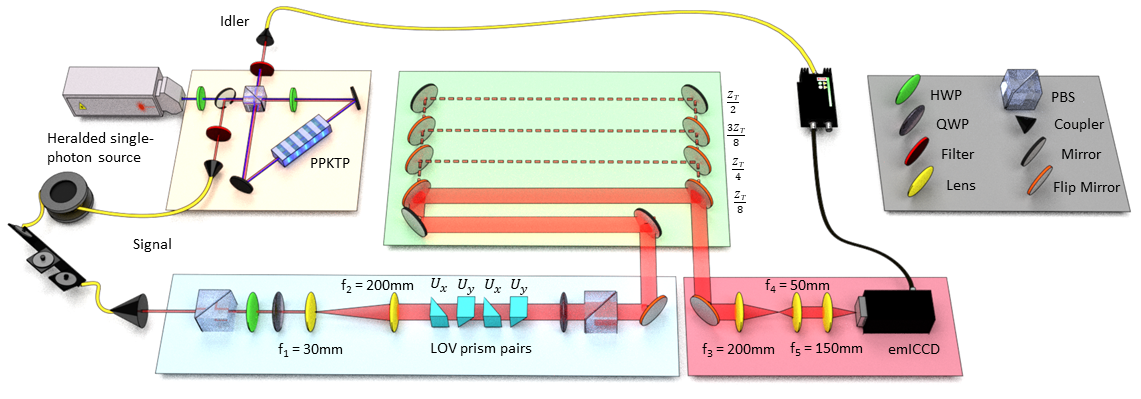}
   \caption{Schematic of the experimental setup. Correlated photon pairs are generated via type-II spontaneous parametric down-conversion in a Sagnac interferometer and coupled into single mode fibers (SMF). A singles rate of $18$~kHz and a coincidence rate of $1.5$~kHz are measured after the SMF. After propagating through a $30$~m long fiber, the signal photon is sent through a telescope with $8.3$x magnification, $N=2$ sets of LOV prism pairs and a polarization filter. The free-space propagation $z$, can be varied via different flip mirror combinations. The signal photons are then imaged onto an intensified electron-multiplying CCD (emICCD), triggered by the detection of the corresponding idler. The imaging arrangement in the detection unit consists of a telescope with $4$x demagnification ($f_3$ and $f_4$ lenses) followed by a single-lens ($f_5$) that images the beam onto the detection plane of the emICCD.
}
   \label{fig:setup}
\end{figure*}

Theory predicts the same self-imaging phenomenon for single photons as well. We describe the free-space propagation of single photons by a complex-valued transverse field distribution $E(x,y)$ convoluted with the Fresnel propagator 
\begin{equation}
K_F(x,y,z) = \frac{e^{i k z}}{i \lambda z}\exp\left[\frac{i k}{2 z} (x^2+y^2)\right],
    \label{eq:FresnelPropagator}
\end{equation}
where $k$ is the wavevector. The field $E(x,y)$ at position $z$ is evaluated via 
\small
\begin{equation}
    E(x,y,z) = \frac{e^{i k z}}{i \lambda z} \iint dx' dy' \, E(x',y',0) \, e^{\frac{ik}{2 z}[(x-x')^2+(y-y')^2]}.
\end{equation}
\normalsize
\figref{fig:TalbotTheo}(a) shows the simulated transverse intensity distribution, $I(x,y)=\vert\langle L\vert\psi^{N=2}_\text{LOV}\rangle\vert^2$, before beam propagation.  \figref{fig:TalbotTheo}(b) and \figref{fig:TalbotTheo}(d) depict the intensity distribution in the $yz$-planes at $x = a/4$ for the initial states $\langle L\vert\psi^{N=2}_\text{LOV}\rangle$ and $\vert\langle L\vert\psi^{N=2}_\text{LOV}\rangle\vert$, respectively. \figref{fig:TalbotTheo}(c) and \figref{fig:TalbotTheo}(e) illustrate the intensity distribution in the $xy$-planes for specific propagation distances. We observe that the initial phase profile defines the transverse intensity pattern at fractional Talbot distances. Furthermore, it can be observed that the OAM phase structure induces an asymmetry between the intensity distributions at propagation distances $\{ z_T/8$,$ z_T/4$,$3z_T/8\}$ and $\{7 z_T/8$, $3 z_T/4$,$5 z_T/8\}$.

The experimental setup is schematically depicted in \figref{fig:setup}. Degenerate photon pairs are prepared using type-II spontaneous parametric down-conversion in a Sagnac interferometer. We pump a $10$~mm long periodically-poled KTP crystal (PPKTP) with a continuous wave diode laser ($404.8$~nm) to produce correlated photon pairs centered at $\lambda_{SP}=810.8$~nm with a spectral bandwidth of $0.4$~nm. With the pump horizontally polarized, we measure the second-order correlation function at zero time delay to be $g^{(2)}(0) = 0.0251 \pm 0.0011$, implying that two-photon events in any coincidence time window are around $1\%$~\cite{Stevens2014}. Note that a diagonal polarized pump would offer the ability to generate a polarization entangled target state, however,  here we herald the signal by means of measuring the idler (see Ref.~\cite{Vermeyden2015} for further details). The outputs of the Sagnac interferometer are coupled into two single-mode fibers, which allow for a distinct separation of signal and idler. The signal photons are sent through a telescope to magnify the beam by a factor of $8.3$, followed by $N=2$ sets of LOV prism pairs. This configuration prepares a lattice of spin-orbit states where one of the polarization states is coupled to $\ell=1$. Higher values of $\ell$ may be achieved by employing a setup with more LOV prism pairs, while negative values of $\ell$ may be achieved by changing the input polarization state~\cite{sarenac2018generation}.

The polarization state of the signal photon is prepared using a manual fiber polarization controller, polarizing beamsplitter (PBS), half wave plate (HWP) and quarter wave plate (QWP). After transmission through the LOV prism pairs, the signal is filtered with respect to left-handed or right-handed circulary polarized light using a QWP. The free-space propagation of the OAM lattice is then analyzed via an arrangement of flip mirrors that effectively change the propagation distance $z$ before measurement. The single photon detection unit consists of a telescope to demagnify the beam by a factor of 4 ($f_3$ and $f_4$ lenses in \figref{fig:setup}) and a gated intensified electron-multiplying CCD (emICCD PI-Max4: 1024 EMB). The telescope is followed by a single lens ($f_5$ lens in \figref{fig:setup}) which images the plane immediately following the telescope. 

The idler is detected by an avalanche photodiode with no polarization filtering, which acts as a trigger for the emICCD to herald the single photon state. We use a $30$~m spool of single-mode fiber to delay the detection of the signal with respect to the idler to accommodate for the delays from the triggering electronics. We set the delay time between the idler and signal photon for each propagation distance $z$ and use the emICCD camera to align the coincidence window of $3$~ns.

In addition to the single photon setup, we couple light from a linearly polarized laser diode (central wavelength $\lambda_{LD} = 813.4$~nm) into the signal channel in order to compare images generated by single photons versus laser diode light. We measure the intensity profile using a conventional CCD camera (Coherent LaserCam-HR II) at the same positions as the single photon images captured by the emICCD.

\begin{figure*}[ht!]
   \centering
   \includegraphics[width=\textwidth]{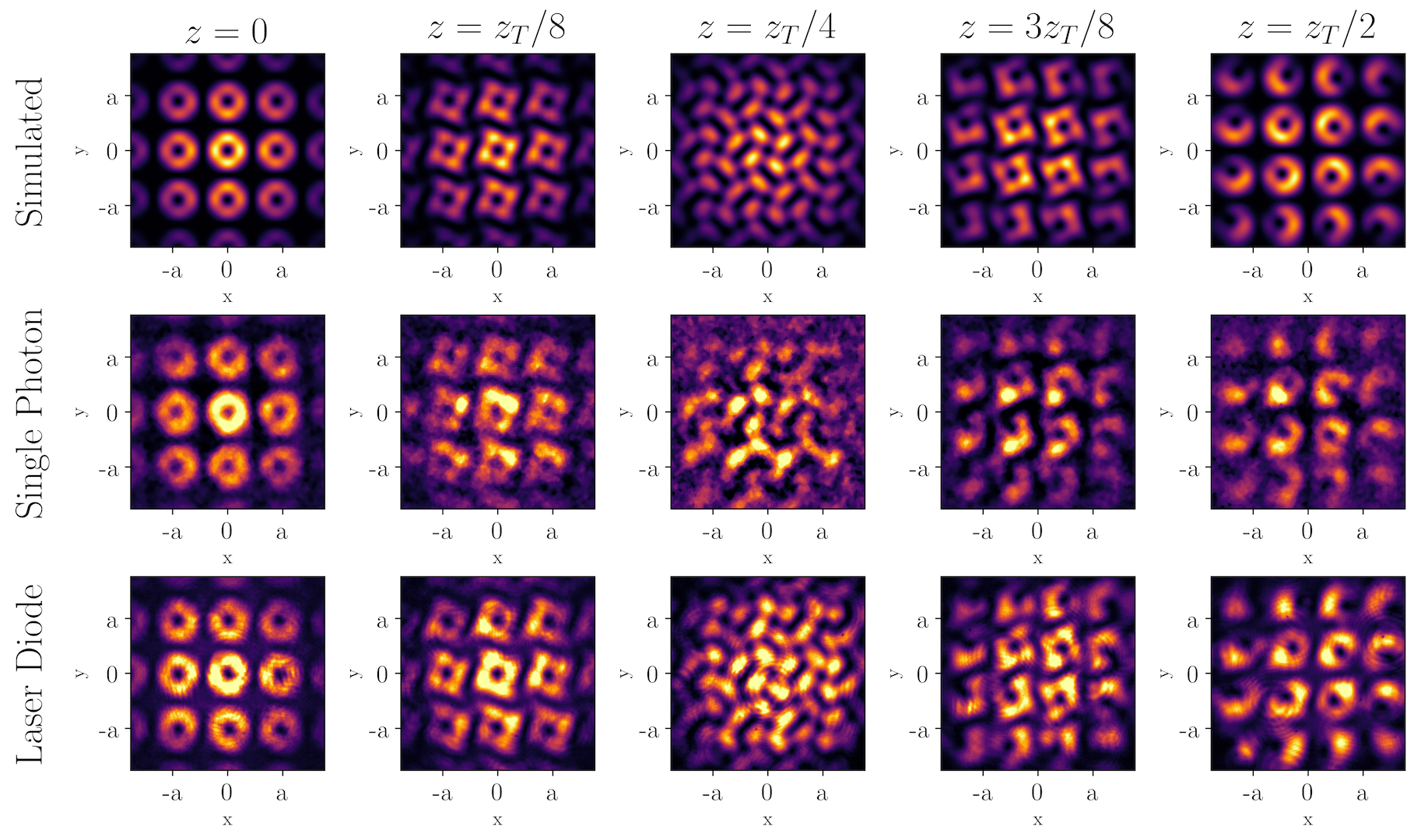}
   \caption{Simulated and experimental self-images at different fractional Talbot lengths. We measure the two-dimensional intensity profile $I(x,y)=\vert \langle L \ket{\Psi^{N=2}_\text{LOV}}\vert^2$ at positions $z\in Z_\text{exp}$. In the simulation, we multiply a Gaussian beam envelope with the same beam waist $w_0$ as in the experiment (i.e., $w_0 = (4.1 \pm 0.05)$~mm) to account for features occurring due to finite beam sizes when propagating along the $z$-axis. For comparison, we couple light from a laser diode into the signal channel, and measure corresponding self-images at the same positions. Good qualitative agreement is found between the simulated and observed profiles. We measure a lattice spacing of $a_{\text{exp}}=(0.573 \pm 0.012)$~mm which corresponds to $(2.229 \pm 0.037)$~mm before the demagnification by the three lens system in the detection unit. From the simulation, we extract a lattice spacing of $a_{\text{sim}}=(0.577 \pm 0.010)$~mm and $(2.301 \pm 0.031)$~mm, respectively.}
   \label{fig:Talbot}
\end{figure*}

\begin{table}[b]
\setlength\tabcolsep{0pt}
\centering
\begin{tabular*}{\columnwidth}{@{\extracolsep{\fill}}cccc}
\toprule
$Z_\text{theo}$ & $Z_\text{exp}$ & 
 \begin{tabular}{@{}c@{}}Measured \\ SNR\end{tabular}
& 
 \begin{tabular}{@{}c@{}}Post-processed \\ SNR \end{tabular}\\
\hline
$0$ & $0.71$~m & $0.584$ & $240.377$  
\\
$z_T/8$ & $2.86$~m & $0.547$ & $181.988$   \\
$z_T/4$ & $4.85$~m & $0.113$ & $102.514$   \\
$3 z_T/8$ & $6.87$~m & $0.159$ & $126.298$   \\
$z_T/2$ & $8.86$~m  & $0.259$ & $264.755$ \\
\hline
\hline
\end{tabular*}
\caption{Experimental propagation distances $Z_\text{exp}$ which correspond to the fractional Talbot distances $Z_\text{theo}$,  and single photon signal-to-noise ratio (SNR). The SNR is given by the ratio of the average signal to the standard deviation of the background. In the third (fourth) column, we list the SNR calculated from raw (post-processed) images.}
\label{tab:SNR}
\end{table}

In \figref{fig:Talbot} we present simulated and measured beam profiles at fractional Talbot distances. Although the theoretical Talbot length is $z_T=16$~m, the propagation distances in the experimental setup were increased by a constant offset of $0.85$~m in order to account for the three lens system in the detection unit~\cite{Goodman2005}. \tabref{tab:SNR} lists the experimental distances, $Z_\text{exp}$, which effectively correspond to the theoretical distances, $Z_\text{theo}$. The diode images were also measured at distances $z \in Z_\text{exp}$. The central wavelength difference of $\vert\lambda_{LD}-\lambda_{SP}\vert=2.6$~nm corresponds to a change in Talbot length $z_T$ of only $\sim 5$~cm. The LOV prisms were realigned in the transverse plane to obtain the most pronounced doughnut structures with the diode laser. 

The observed intensity profiles are measured with a total exposure time of about 1 hour and are processed using background subtraction and an adaptive two-dimensional Gaussian image filter. Including the quadratic phase profiles of the imaging lenses in the simulation yields good agreement between theoretical and observed lattice spacing. For instance, in the case of single photons, we extract from the transverse intensity distribution at $z=0.071$~m a separation between two nearest-neighbor lattice sites of $a_{\text{exp}}=(0.573 \pm 0.012)$~mm from experimental data and $a_{\text{sim}}=(0.577 \pm 0.010)$~mm from the simulation. Additionally, at half Talbot distance $z_T/2$, the expected half period shift $\Delta a$ can be evaluated by comparing the effective pixel positions of the lattice sites at propagation distance $z=0.071$~m with the pixel positions at $z=z_T/2$ yielding $\Delta a_{\text{exp}}=(0.273 \pm 0.015)$~mm and $\Delta a_{\text{sim}}=(0.279 \pm 0.014)$~mm, respectively.

The robustness of the Talbot Effect with a lattice of OAM states is demonstrated by the good qualitative agreement between simulation, single photon, and diode laser images in \figref{fig:Talbot}. The SNR decreases with larger distances, but is increased again depending on the intensity pattern complexity. In \tabref{tab:SNR}, we present the SNR before and after the imaging post-processing for different propagation distances. However, it can be noted that the self-imaging property of this beam can be seen clearly in the similarity between images taken at distances $z = \{0,  z_T/2\}$, with the correct spatial shift. Images at $z = \{z_T/8,  3 z_T/8\}$ show an orientation about each lattice site that appears counter-clockwise in $z_T/8$ and clockwise in $3 z_T/8$. These features are indicative of the OAM state in each lattice site, as shown in \figref{fig:TalbotTheo}(c). Furthermore, gaps in the outermost rings of the $z_T/2$ image can be mitigated by using a beam containing more lattice sites. 

In this work we demonstrated and analyzed the Talbot Effect with single photons prepared in a lattice of OAM states. Heralded single photons are sent through $N=2$ sets of LOV prism pairs and their transverse two-dimensional intensity distribution are measured at various fractional Talbot lengths. The propagation of structured wavefronts is calculated in the near-field and shows good agreement with experimental results. We observe that the initial phase profile defines the transverse intensity pattern at various propagation distances, and thus the Talbot carpet. Future work will scrutinize the connection between OAM and Talbot physics as a new characterization tool. Another avenue of exploration includes the addition of path entangled OAM lattices using the Talbot Effect and the OAM degree of freedom. Other avenues of exploration include the addition of path entangled OAM lattices and the implementation of quantum logic using the Talbot Effect and the OAM degree of freedom. 

This research was supported in part by the Canadian Excellence Research Chairs (CERC) program, the Natural Sciences and Engineering Research Council of Canada (NSERC), the NSERC Discovery program, Canada Research Chairs, Industry Canada and the Canada Foundation for Innovation (CFI), Ontario Research Fund (ORF), and the Canada First Research Excellence Fund (CFREF).

\bibliography{TalbotEffect}

\end{document}